\documentclass[aps,prl,twocolumn,showpacs,superscriptaddress,groupedaddress,nofootinbib]{revtex4}

\usepackage{amsmath}
\usepackage{amssymb}
\usepackage{amsthm}
\usepackage{bm}
\usepackage{dcolumn}
\usepackage{float}
\usepackage{graphicx}
\usepackage{subfigure}

\begin{document}

\title{Magnetar Asteroseismology with Long-Term Gravitational Waves}
\author{Kazumi Kashiyama}
\affiliation{Department of Physics,~Kyoto~University,~Kyoto~606-8502,~Japan}

\author{Kunihito Ioka}
\affiliation{Theory Center,~KEK(High~Energy~Accelerator~Research~Organization),~Tsukuba~305-0801,~Japan}

\begin{abstract}
Magnetic flares and induced oscillations of magnetars (supermagnetized neutron stars) are promising sources of gravitational waves (GWs). We suggest that the GW emission, if any, would last longer than the observed x-ray quasiperiodic oscillations (X-QPOs), calling for the longer-term GW analyses lasting a day to months, compared to than current searches lasting. Like the pulsar timing, the oscillation frequency would also evolve with time because of the decay or reconfiguration of the magnetic field, which is crucial for the GW detection. With the observed GW frequency and its time-derivatives, we can probe the interior magnetic field strength of $\sim 10^{16}$ G and its evolution to open a new GW asteroseismology with the next generation interferometers like advanced LIGO, advanced VIRGO, LCGT and ET.
\end{abstract}
\pacs{95.85.Sz,97.60.Jd}
\maketitle

\vspace{2mm}
In the upcoming years, the gravitational wave (GW) astronomy will be started by the 2nd generation GW detectors~\cite{Harry:2010,VIRGO,Kuroda:2010}, such as advanced LIGO, advanced VIRGO and LCGT, and the 3rd generation ones like ET~\cite{Hild:2010} in the $10$ Hz-kHz band. These GW interferometers will bring about a greater synergy among multimessenger (electromagnetic, neutrino, cosmic ray and GW) signals. One of the most important targets is the neutron star (NS) oscillations. We can infer the NS interior from eigenfrequencies of the waveform, and could open {\it gravitational wave asteroseismology}~\cite{Andersson:1998,Kokkotas:1999}.

Highly magnetized NS's with $\sim 10^{14\mbox{-}16} \mathrm{G}$ so-called magnetars~\cite{Thompson:1992,Thompson:1995,Duncan:1998} are promising candidates for the GW asteroseismology. They are observed as soft gamma-ray repeaters~(SGRs) and anomalous x-ray pulsars~(AXPs) in our galaxy and the LMC, whose emission is thought to be powered by the dissipation of magnetic energy. SGRs emit repeated bursts and sometimes giant flares~(GFs) with huge energy $\sim 10^{44-46}$ erg \cite{Mazets:1979,Hurley:1999,Hurley:2005}.
In order to prove that the energy source is the magnetic field, 
not the mass accretion \cite{Chatterjee:1999iy},
the GW-scopy of the internal field is indispensable.

The magnetar GFs would excite the seismic oscillations 
with GWs~\cite{Abadie:2010,Ioka:2001,Corsi:2011}.
The GFs are thought to be produced by 
a release of the accumulated magnetic stress in the crust or inner region.
A part of the GF energy will tap the stellar oscillations,
especially polar modes that vary the moment of inertia and emit GWs
\cite{Sotani:2009}.
At least the reduction of the magnetic stress would change
the stellar deformation and begin polar oscillations around a new equilibrium.

Actually, the quasiperiodic oscillations with $\sim 100$ Hz have been discovered in the x-ray tails of GFs~(X-QPOs)~\cite{Israel:2005} and recently 
activated the searches of GWs from magnetar GFs. The GW energy within the X-QPO duration $\sim 200$ sec has been limited below $\sim 10\%$ of the electromagnetic energy~\cite{Abadie:2010}.
X-QPOs can be interpreted by the axial oscillations propagating
to the magnetospheric emission 
via magnetic fields~\cite{Israel:2005,Sotani:2008,Umin:2010}.

In this Letter, we suggest that the GWs from magnetar GFs would last much longer than the X-QPOs if 
the GW frequencies are close to the X-QPO frequencies [see Eq.(\ref{eq:tauGW})].
We show that the long-term GW analyses from a day to months are necessary to detect the GW, even if the GW energy is comparable to the electromagnetic energy. The current searches of $\lesssim 200$ sec are not long enough; this is because the GW luminosity is proportional to $\text{(frequency)}^6$. Thus, by fixing the GW energy budgets, $\sim 100 \mathrm{Hz}$ GWs could persist up to $10^6$ times longer than $\sim \mathrm{kHz}$ GWs of typical fluid modes like p-modes.

In most previous studies,
the fluid modes were employed for GWs.
However, $\sim 100$ Hz oscillations rather than 
$\sim \mathrm{kHz}$ fluid modes will be excited more effectively
in the magnetar GFs
because the initial spike of GF had a rise time $\sim 0.01$ sec~\cite{Mazets:1979,Hurley:1999,Hurley:2005}
and its inverse $\sim 100$ Hz is most resonant.
Actually X-QPOs have this frequency.
This is also about the internal Alfv$\acute{\text{e}}$n crossing frequency [see Eq.(\ref{eq:fa})], which is one of the reasons to believe the magnetic origin of GFs \cite{Thompson:1995}.

So, we propose that GWs from magnetar GFs, if any, are largely produced by
the polar Alfv$\acute{\text{e}}$n oscillations,
which last longer than the axial-type X-QPOs.
This opens an interesting possibility to
directly measure the internal magnetic field 
from the observed frequency $f_a \propto B$
since the restoring force is magnetic
and the polar Alfv$\acute{\text{e}}$n modes have discrete frequencies
\cite{Sotani:2009}.

The long-term GW emission also makes it possible to directly measure the magnetic field decay or reconfiguration inside the NS, because the mode frequency $f$ depends on $B$ that evolves with time. The Taylor expanded phase of the GW,
\begin{equation}\label{Taylar}
\Psi(t) = \psi_0 + 2 \pi \left\{ f (t-t_0) - \frac{1}{2}\dot{f} (t-t_0)^2 + ... \right\},
\end{equation}
has the first frequency derivative~(1FD) term and also the higher frequency derivative terms~(2FD and so on) like the pulsar timing search. The FD terms affect the signal to noise ratio (S/N) even if we assume the minimum magnetic field decay required to supply energy for the observed emission from magnetars [see Eq.(\ref{f_a_dot})].

In our picture, GWs are produced by the polar oscillations
while X-QPOs are caused by the axial ones.
This is implied by the amplitude of X-QPOs $\sim 0.1$~\cite{Israel:2005},
which is much larger than 
the expected polar oscillation amplitude $\lesssim 10^{-4}$ 
[see Eq.~(\ref{eq:amp})]. 
Thus, the GW searches beyond the X-QPO duration are meaningful.

To clarify the above statement, we first make an order-of-magnitude estimate before the detailed analysis. We take the fiducial minimum magnetic field inside the NS as $B = 10^{16} \mathrm{G}$ because the total magnetic energy $\sim (B^2/8\pi) \cdot (4\pi R^3/3) \sim 10^{49}$ erg should supply $\sim 10^{46} \mathrm{erg} \times 10^4 \mathrm{yr} / 10^2 \mathrm{yr} \sim 10^{48} \mathrm{erg}$ for the GF activities once per century (with three GFs out of $\sim 10$ sources within 30 years) during the lifetime of magnetars, $\tau_{mag} \sim 10^4 \mathrm{yr}$, and the ionospheric observations suggest that $\sim 9$ times energy is released outside the X-$\gamma$ ray band~\cite{Inan:1999}.

When the interior magnetic fields are deformed by the GFs, the induced difference in the moment of inertia of the star $\epsilon$ is given by
\begin{equation}\label{eq:amp}
\frac{\Delta I}{I} \equiv \epsilon \lesssim 
\epsilon_{\max} \equiv \frac{(B^2/8\pi ) \cdot (4\pi R^3/3) }{GM^2/R} \sim 3 \times 10^{-5} B_{16}{}^2,
\end{equation}
where $I \approx 2MR^2/5$ is the moment of inertia of the magnetar, $B_{16} \equiv (B/10^{16} \mathrm{G})$ is the interior magnetic field, and $M$, $R$ are the mass and the radius, respectively~\cite{Ioka:2001,Ioka:2003nh}. Hereafter, we fix $R = 10^6 \mathrm{cm}$ and $M = 1.4 M_{\odot}$. 
The oscillation energy or the GW energy is about the gravitational energy shift caused by the stellar deformation, 
\begin{equation}\label{eq:E}
E_{GW} \approx (\epsilon^2/5 )\cdot (GM^2/R) \lesssim 8 \times 10^{43} B_{16}{}^4 \ [\mathrm{erg}].
\end{equation}
The oscillation frequencies are determined by
the Alfv$\acute{\text{e}}$n sound crossing,
\begin{equation}\label{eq:fa}
f_a = \frac{v_a}{R} = \frac{1}{R} \cdot \frac{B}{\sqrt{4 \pi \rho}} \sim 100 B_{16} \ [\mathrm{Hz}], 
\end{equation}
where $v_a$ is the Alfv$\acute{\text{e}}$n velocity, and $\rho \approx M/(4\pi R^3/3)$ is the mass density. If these oscillations are polar mode~\cite{Sotani:2009}, in which the moment of inertia changes, GWs are emitted. The luminosity of the GWs are estimated to be 
\begin{eqnarray}\label{eq:L}
L_{GW} & \approx & (G/5c^5) \cdot \langle \dddot{{I\hspace{-.65em}-}}_{ij} \dddot{I\hspace{-.65em}-}_{ij} \rangle \approx (G/5c^5) \cdot ( \epsilon I (2 \pi f_a)^3)^2 \notag \\ &\lesssim& 2 \times 10^{37} B_{16}{}^{10} \ [\mathrm{erg/sec}].
\end{eqnarray}
Then the GW duration after the GF becomes
\begin{equation}\label{eq:tauGW}
\tau_{GW} \approx E_{GW}/L_{GW} \sim 4 \times 10^{6} B_{16}{}^{-6} \ [\mathrm{sec}].
\end{equation}
The characteristic GW amplitude $h_c$~\cite{Thorne:1987} is given by
\begin{eqnarray}\label{h_c}
h_c &\approx & \left( \frac{G L_{GW}}{\pi^2 c^3 d^2 f_a^2}\right)^{1/2} \cdot \sqrt{f_a \tau_{GW}} \notag \\ & \lesssim & 5 \times 10^{-23} B_{16}{}^{3/2} d_{10}{}^{-1}, 
\end{eqnarray}
where $d_{10} = d/10 \mathrm{kpc}$ is the distance from the source. Therefore, if the GW energy is $\gtrsim 10^{44} \mathrm{erg}$ (only a fraction of the GF energy $\sim 10^{44-46}$ erg) and the data taking lasts for a day to months, the GWs from magnetar GFs within $10~\mathrm{kpc}$ are detectable with enough S/N by the next generation detectors.
The event rate will be $\sim 1/10\mathrm{yr}$, and more for weaker flares.
We should note that the GW energy or the deformation $\epsilon$
has a large uncertainty.
The minimum deformation would be $\epsilon \sim \epsilon_{\max}/100$ in Eq.(\ref{eq:amp})
since $\sim 1\%$ of magnetic energy is released in a GF.
We can rescale the following results with 
${\rm S/N} \propto h_c \propto \epsilon$.
Meanwhile, Eqs.(\ref{eq:fa}) and (\ref{eq:tauGW}) do not depend on $\epsilon$, and thus are relatively robust. 

Since the magnetic fields of magnetars are considered to decay significantly in the lifetime $\tau_{mag} \sim 10^4 \mathrm{yr}$, the magnetic field are decaying at least at the rate
\begin{equation}
\dot{B} \approx -B/\tau_{mag} \sim -10^{5} B_{16} \ [\mathrm{G/sec}].
\end{equation}
This leads to the evolution of the Alfv$\acute{\text{e}}$n frequency 
\begin{equation}\label{f_a_dot}
\dot{f_a} \approx \frac{1}{R} \cdot \frac{\dot{B}}{\sqrt{4 \pi \rho}} \sim -10^{-9} B_{16} \ [\mathrm{Hz/sec}]. 
\end{equation}
During the GW emission, the oscillation frequency changes by 
\begin{equation}
|\Delta f_a| \sim |\dot{f_a} \times \tau_{GW}| \sim 4 \times 10^{-3} B_{16}{}^{-5} \mathrm{Hz}.
\end{equation}
In principle, the frequency resolution of GW is given by the inverse of the observational time $T_{obs}$,
\begin{equation}
\Delta f_{obs} \sim T_{obs}{}^{-1} \sim \tau_{GW}{}^{-1} \sim 3\times 10^{-7} B_{16}{}^6 \mathrm{Hz}
\end{equation}
for long enough data taking. Therefore, we could resolve the GW frequency evolution to prove the magnetic field decay rate inside the magnetar for $B_{16} \lesssim 3.0$. In other words, the 1FD term could be crucial for keeping S/N in the GW detection.
If the magnetic field does not decay constantly in the GF phase, higher FD terms may be important, where the long-term frequency-time analysis~\cite{Thrane:2010} may be useful.

Next, we shall move on to more details with the matched filtering Fisher analyses. We only consider the $l=2, m=0$ mode as the magnetar oscillation because one-sided outflow has been confirmed by the radio observation of the GF from SGR 1806-20 ~\cite{Taylor:2005}. This suggests that a GF is a pointlike energy release from a NS, which would mainly trigger the $l=1, m=0$ plus $l=2, m=0$ mode oscillations. We take $m=0$ since there is no specific direction along the azimuthal angle and the rotation is slow ($\sim 0.1$ Hz) compared with the Alfv$\acute{\text{e}}$n oscillations. The $l=1, m=0$ mode is the dipole oscillation which does not contribute to the GW emission.

\begin{figure}[h]
\begin{center}
\includegraphics[width=50mm, angle=-90]{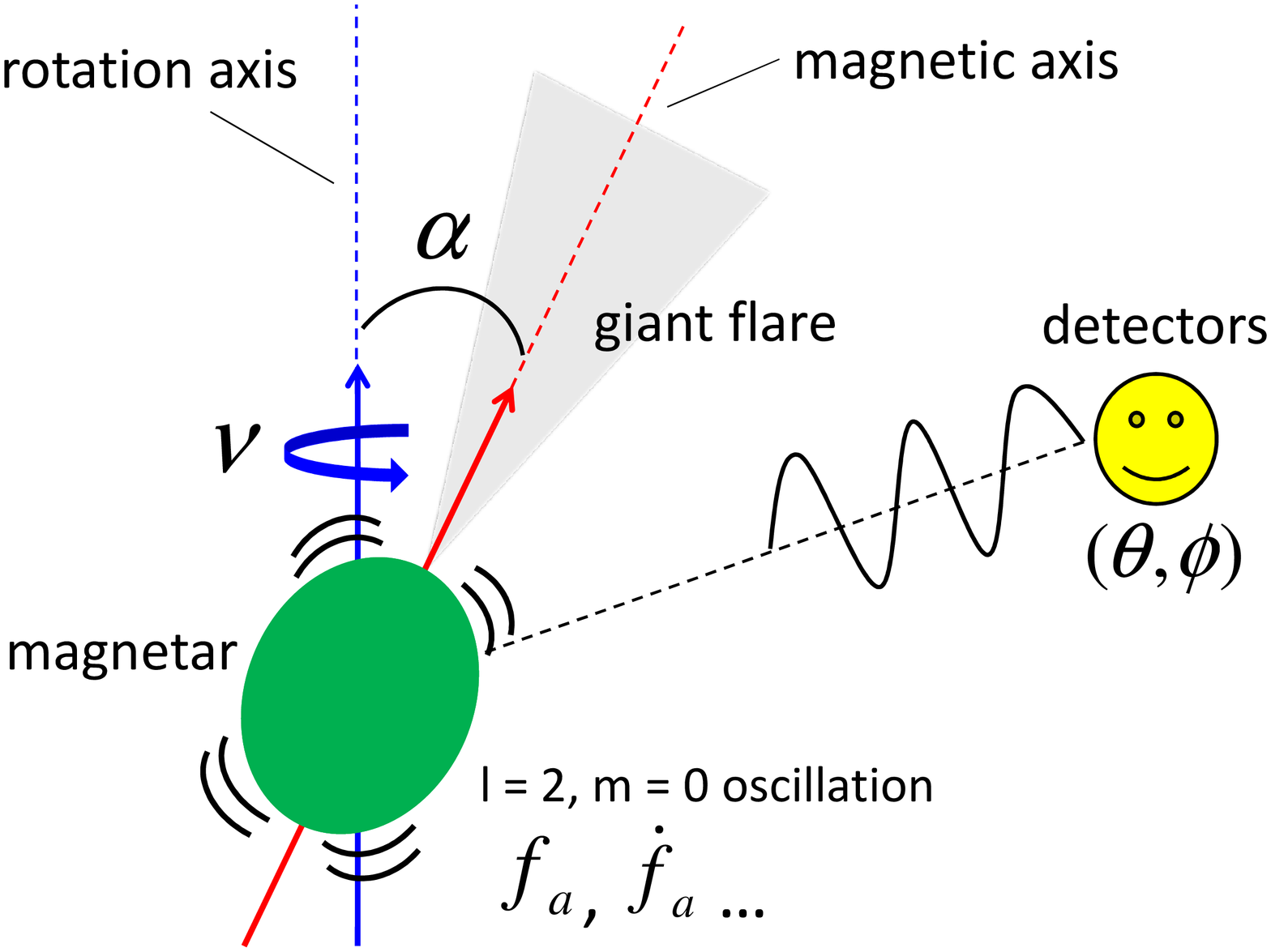}
\caption{The configuration of the magnetar and detectors.}\label{oci_and_rot}
\end{center}
\end{figure}

Taking a misalignment between the polar oscillation and the rotation axes as in Fig.\ref{oci_and_rot},
the waveform of the observed plus~(+) and cross~($\times$) GWs are 
obtained from the quadrupole formula,
\begin{equation}
h_{+} + i h_{\times} = \frac{\epsilon I}{d} ({\cal A} + i {\cal B} ) \cos\Psi(t)  \exp\left( -\frac{t}{\tau_{GW}} \right),
\end{equation}
where
\begin{eqnarray}
{\cal A} = \frac{2}{\sqrt{3}} \left( \frac{d\Psi}{dt} \right)^2 \left\{ (\cos^2 \theta +1)\sin^2 \alpha \cos2(2\pi\nu t+\phi) \right. \notag \\ \left. + \sin^2\theta(3\cos^2\alpha-1)+\sin2\theta \sin2\alpha \cos(2\pi\nu t +\phi) \right\},
\end{eqnarray}
\begin{eqnarray}
{\cal B} = \frac{2}{\sqrt{3}} \left( \frac{d\Psi}{dt} \right)^2 \left\{ \cos\theta \sin\alpha \sin2(2\pi\nu t+\phi) \right. \notag \\ \left. + \sin\theta \sin2\alpha \sin(2\pi\nu t +\phi) \right\}.
\end{eqnarray}
The stellar oscillation phase $\Psi(t)$ can be Taylor expanded as Eq.(\ref{Taylar}). In this Letter, we stop the expansion at the 1FD term, and add $\psi_0$, $f_a$ and $\dot{f_a}$ as the parameters for the GW waveform. We neglect the proper motion of the magnetar. As a whole, the parameters are the strength of deformation $\epsilon$, the mean moment of inertia $I$, the distance to the magnetar $d$, the rotational frequency of the magnetar $\nu$, the angle between the magnetar rotation and the oscillation axes $\alpha$, the frequency of the oscillation $f_a$, its evolution rate $\dot{f_a}$, the initial phase of the oscillation $\psi_0$, the duration of the GW emission $\tau_{GW}$, and the observational polar and azimuthal angle $\theta$ and $\phi$. The position $(\theta,\phi)$ can be determined independently by the x-ray or radio observations. Here, $I$, $d$ and $\epsilon$ are completely degenerated. We fix $I = 10^{45} \mathrm{erg\cdot sec^2}$, $d = 10 \mathrm{kpc}$ and leave $\epsilon$ as a free parameter. According to Eqs.(\ref{eq:amp}),(\ref{eq:fa}),(\ref{eq:tauGW}) and (\ref{f_a_dot}), we set $f_a = 100 B_{16} \mathrm{Hz}$, $\dot{f_a} = -10^{-9} B_{16} \mathrm{Hz/sec}$, $\tau_{GW}= 4 \times 10^6 B_{16}{}^{-5} \mathrm{sec}$ and $\epsilon = 3 \times 10^{-5} B_{16}{}^2$ (i.e., optimal case).
We also set $\alpha = \pi/4$, $\nu = 0.1 \mathrm{Hz}$, $\theta = \pi/2$, $\phi = \pi/4$ and $\psi_0 = \pi/4$. 

In the matched filtering analysis, we can compute the determination accuracy of these parameters using the Fisher matrix formalism with a waveform template in the frequency domain $\tilde{h}(f)$~\cite{Finn:1992, Cutler:1998}. The variance-covariance matrix of the parameter estimation error $\Delta \gamma_i$ is given by the inverse of the Fisher information matrix $\Gamma_{ij}$ as $\langle \Delta\gamma_i \Delta\gamma_j \rangle = (\Gamma^{-1})_{ij}$. The Fisher matrix becomes
\begin{equation}\label{fisher_1}
\Gamma_{ij} = 4 \mathrm{Re} \sum_{\alpha = +,\times} \int \frac{df}{S_n(f)} \frac{\partial \tilde{h}_{\alpha}^*(f)}{\partial \gamma_i}\frac{\partial \tilde{h}_{\alpha}(f)}{\partial \gamma_j},
\end{equation}
where $S_n(f)$ is the noise spectrum. The accuracies of the interested parameters are described as $\Delta B/ B = \Delta f_a/f_a$, $\Delta \dot{B}/ \dot{B} = \Delta \dot{f}_a/\dot{f}_a$ and $\Delta E_{GW}/ E_{GW} = 2\Delta \epsilon/\epsilon$. The resolution for the position of the magnetar is 
\begin{equation}
\Delta\Omega = 2\pi \sqrt{\langle \Delta \mu^2 \rangle \langle \Delta \phi^2 \rangle - \langle \Delta \mu \Delta \phi \rangle^2},
\end{equation}
where $\mu = \cos\theta$. The S/N is given by 
\begin{equation}\label{fisher_2}
(\mathrm{S/N})^2 = 4\sum_{\alpha = +,\times} \int \frac{df}{S_n(f)} |\tilde{h}_{\alpha}(f)|^2.
\end{equation}
We integrate the waveform in Eqs.(\ref{fisher_1}) and (\ref{fisher_2}) for the duration $T_{obs}$. We refer advanced LIGO~\cite{Harry:2010}~(2nd generation) and ET~\cite{Hild:2010}~(3rd generation) for a detector noise spectrum $S_n(f)$.

\begin{table}[htbp]
\caption{S/N and parameter determination accuracy \\ for the optimal case ($\epsilon = \epsilon_{\mathrm{max}}$)}.
\label{results}
{\renewcommand\arraystretch{1.5}
\begin{ruledtabular}
\begin{tabular}{c|cc|cc}
detector & \multicolumn{2}{c|}{2nd generation} & \multicolumn{2}{c}{3rd generation} \\ \hline 
$B_{16}$ & $1.0$ & $2.0$ & $1.0$ & $2.0$ \\ 
$T_{obs}(\gtrsim \tau_{GW})$ & 4 months & 2 days & 4 months & 2 days \\ \hline
S/N & $11$ & $23$ & $170$ & $350$\\ 
$\Delta B / B $ & $6.4 \times 10^{-10} $ & $1.0 \times 10^{-8}$ & $4.0 \times 10^{-11}$ & $6.5 \times 10^{-10}$\\
$\Delta\dot{B} / \dot{B} $ & $8.9 \times 10^{-6}$ & $9.0 \times 10^{-3}$ & $5.6 \times 10^{-7}$ & $5.8 \times 10^{-4}$\\  
$\Delta\nu / \nu$ & $4.2 \times 10^{-8}$ & $1.3 \times 10^{-6}$ & $2.7 \times 10^{-9}$ & $8.4 \times 10^{-8}$\\ 
$\Delta E_{GW} / E_{GW}$ & $0.27$ & $0.14$ & $1.7 \times 10^{-2}$ & $8.8 \times 10^{-3}$\\ 
$\Delta \tau_{GW} / \tau_{GW}$ & $0.18$ & $0.090$ & $1.2 \times 10^{-2}$ & $5.8 \times 10^{-3}$\\ 
$\Delta\alpha / \alpha$ & $0.030$ & $0.015$ & $1.9 \times 10^{-3}$ & $9.8 \times 10^{-4}$\\ 
$\Delta\Omega$ & $0.044$ & $0.010$ & $1.7 \times 10^{-4}$ & $4.2 \times 10^{-5}$
\end{tabular}
\end{ruledtabular}
}
\end{table}

Table.\ref{results} shows the S/N and the parameter determination accuracy assuming enough data taking, $T_{obs} \gtrsim \tau_{GW}$. For the optimal case ($\epsilon = \epsilon_{\mathrm{max}}$), even 2nd generation detectors can detect the GWs with enough S/N. The detection of the GW is possible when $\epsilon \gtrsim 0.1\epsilon_{\mathrm{max}}$ and $\epsilon \gtrsim 0.01\epsilon_{\mathrm{max}}$ for 2nd and 3rd generation detectors, respectively. Thanks to the large Q value $\sim f_a\times \tau_{GW} \sim 10^8$, parameters related to the phase of the GW~(i.e., $f_a$, $\dot{f}_a$ and $\nu$) could be determined accurately. When fixing $B$, errors scales as $(\mathrm{S/N})^{-2}$ for $\Delta \Omega$, and $(\mathrm{S/N})^{-1}$ for other parameters. 

\begin{figure}[htbp]
\includegraphics[width=85mm]{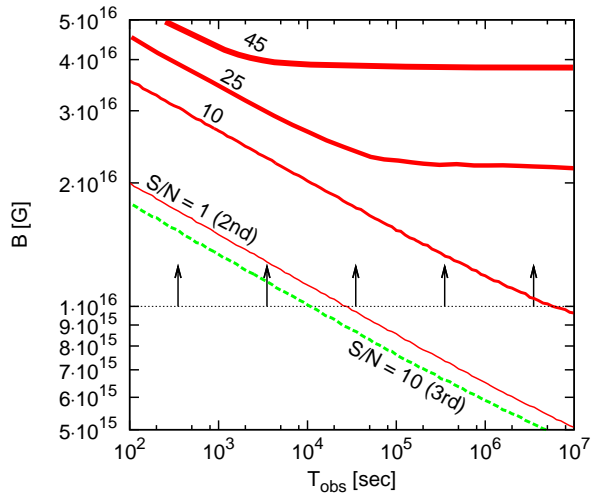}
\caption{The contour lines of S/N in the plane of the magnetic field of magnetars and the GW observation time. Solid~(dashed) lines show S/N $= 1,10,25,45$~(S/N $= 10$) using a 2nd~(3rd) generation detector. $B = 10^{16} \mathrm{G}$ is the minimum interior magnetic field required for magnetar activities.}
\label{SN}
\end{figure}
\begin{figure}[htbp]
\includegraphics[width=85mm]{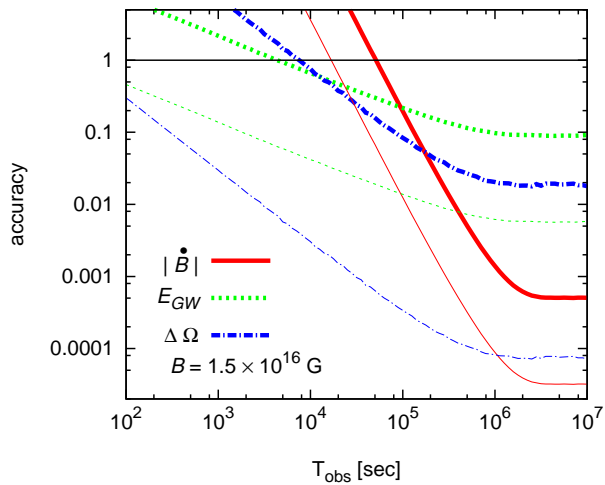}
\caption{The parameter determination accuracies with respect to the GW data taking time. The solid, dotted and dotted-dash lines show $|\dot{B}|$, $E_{GW}$ and $\Delta \Omega$, respectively. Thin~(thick) lines shows the results using 2nd~(3rd) generation detectors. We set $B = 1.5\times 10^{16} \mathrm{G}$.}
\label{Accuracy}
\end{figure}
Figs.\ref{SN} and \ref{Accuracy} show the dependence of the S/N and the accuracies on the data taking time $T_{obs}$. We can clearly find that the long-term data taking for a day~($\sim 10^5 \mathrm{sec}$) to months~($\sim 10^7 \mathrm{sec}$) is necessary for high S/N and parameter accuracies, especially $\dot{B}$.

Since a NS has various oscillation modes, it is hard, at this stage, to answer which oscillation mode is mainly excited at the GF. Also, we should consider the mode coupling between the polar Alfv$\acute{\text{e}}$n mode and other modes. If the coupling is strong, the polar Alfv$\acute{\text{e}}$n mode oscillation cannot last so long as $\sim$ a day to months even if the GF energy are initially injected to the polar Alfv$\acute{\text{e}}$n mode, and the dependence of Alfv$\acute{\text{e}}$n mode frequencies on the magnetic field may become different from Eq.(\ref{eq:fa}). The effects of the crust and superfluidity are also to be investigated in the future.

The change in the rotational frequency $\dot{\nu}$ could become important as the 1FD. Although the magnetar spindown rate is usually small $\sim -10^{-12} \mathrm{Hz/sec}$ compared with Eq.(\ref{f_a_dot}), the sudden increase in the GF phase $\Delta \nu / \nu \sim -10^{-4}$ has been reported in the August 27, 1998 event from SGR 1900+14~\cite{Woods:1999}. Since $\nu \ne f_a$, we could also measure $\dot{\nu}$ independently.

Finally, we should mention the strategic change in the analysis of the GW waveform. When discussing GWs from a single NS or black hole, the waveform usually has a short damping time $\lesssim 10 \mathrm{sec}$~\cite{Tsunesada:2005,Abbott:2009}, hence the evolution of the waveform due to the daily and yearly motion of the earth could be neglected. However in the case of the long-term oscillations with $\gtrsim$ a day, we have to take into account these effects as in the analysis of GW from a pulsar~\cite{Neibauer:1993,Akutsu:2008}.

We thank N. Seto, M. Ando, K. Yagi and T. Nakamura for much useful advice. This work was supported by the Grant-in-Aid for the Global COE Program "The Next Generation of Physics, Spun from Universality and Emergence" from the Ministry of Education, Culture, Sports, Science and Technology (MEXT) of Japan and KAKENHI, 19047004, 21684014, 22244019, 22244030.

\end{document}